\documentclass[twocolumn,showpacs,preprintnumbers,amsmath,amssymb]{revtex4}
\usepackage{amsmath,amssymb,graphics,epsfig,subfigure,color}
\usepackage{color}

\begin{document}
\newcommand {\nn}    {\nonumber}

\title{Deformed brane with finite extra dimension}
\author{Yu-Xiao Liu\footnote{liuyx@lzu.edu.cn},
        Chun-E Fu\footnote{fuche08@lzu.edu.cn, corresponding author},
        Heng Guo\footnote{guoh06@lzu.edu.cn},
        Hai-Tao Li\footnote{liht07@lzu.edu.cn}}.
 \affiliation{Institute of Theoretical Physics,
              Lanzhou University, Lanzhou 730000,
             People's Republic of China}


\begin{abstract}
We construct a deformed brane solution generated by a double-kink scalar field and a dilaton scalar field. In this brane scenario the extra dimension is finite, which is due to the introduction of the dilaton field with special form. The finity of the extra dimension will result in the localization of the zero mode for the vector fields. While the localization of the Kalb-Ramond fields {depends} on the coupling to the dilaton. For the fermion fields, {with different values of the dilaton-fermion coupling constant}, there are three types of the effective potential for the fermion KK modes. Moreover, we investigate the effect of the deformation of the brane on the localization, and find that the number of the resonances will increase with the distances of the two sub-branes.

\end{abstract}



\maketitle

\section{Introduction}

Extra dimensions and braneworld theories \cite{Rubakov1983,Akama,Antoniadis} have attracted more and more attention, because { they} can provide an interesting new way for solving problems involving cosmological constant and hierarchy \cite{ADD,RS,Neupane}, such as the Arkani-Hamed-Dimopoulos-Dvali (ADD) model and the the Randall-Sundrum (RS) model. In these two models, the extra dimensions are flat and compact for the former one and infinite for the latter one, but both the branes in the two models have no thickness. So the thick brane models were built, which are naturally {generated} by one or more real scalar fields
\cite{DeWolfe,GremmPLB2000,ThickBrane1,ThickBraneWeyl,varios,SingletonFermions,1006.4240,
Guerrero2002,ThickBraneDzhunushaliev,ThickBraneBazeia,ShtanovJCAP2009,
Bazeia0808,two-fieldbrane,two-scalarbrane1,
BranewithTwoScalars,YZhong2010}, and the extra dimensions are usually infinite in these models.

In Ref.~\cite{PRL141602}, the author investigated the splitting of the thick Minkowski brane {generated} by a complex scalar field coupled to gravity, and showed that the deformation of the brane is due to a first-order phase transition. And in another aspect, the authors in Refs.~\cite{deformed1,BlochBrane,zhao2011jhep,zhaoliu2010} built brane models described by real or complex scalar fields, which also engender internal structure. The internal structure depends on the properties of the scalar fields.

In this paper we would like to construct a deformed brane with finite extra dimension and study the effect of the deformation on the localization of various matter fields. Usually, a scalar potential with two vacua would generate a single thick brane. Such potential could be chosen as a usual $\phi^4$ one. However, in order to generate a deformed brane containing two sub-branes, it needs a potential with three vacua. So for simplicity, the scalar potential is selected as the usual $\phi^6$ one for the purpose of generating the splitting brane. On the other hand, the extra dimension for the one-scalar generated brane is infinite, which is the reason that the vector can not be localized on the brane in 5-dimensions. So in the paper we add another scalar, a dilaton, with which the extra dimension would become finite and the vector could be localized on the brane. In a word, we would obtain a thick flat split-brane solution with internal structure and the extra dimension is finite in this brane model. Then we will investigate the localization of various bulk fields on this brane world.

The localization of various bulk matter fields is important to build up the standard model. It has been known that massless scalar fields and gravitons can be localized on branes with exponentially decreasing warp factor \cite{RS,BajcPLB2000}. The spin-1 Abelian vector fields usually can only be localized on the RS brane in some higher-dimensional cases \cite{OdaPLB2000113}, or on the thick de Sitter brane and the Weyl thick brane \cite{LiuJCAP2009,Liu0803}. But in our thick flat brane model, the vector fields also can be localized, which can be seen in the following discussion.

The antisymmetric Kalb-Ramond (KR) tensor field $B_{\mu\nu}$ was first introduced in the string theory as a massless mode. Then it was used to explain the torsion of the space-time in the Einstein-Cartan theory. While in theories of extra dimensions, they indicate new types of particles~\cite{duality}. So any observational effect involving the KR fields is a window into the inaccessible world of very high energy physics. Thus the investigation of KR fields in the context of extra dimension theories has been carried out~\cite{branep1,KRfieldinRS1,KRfieldinRS2,KRthickbrane,KR1006.1366,KRfieldinRS3}. We are also interested in the localization of the KR field.

The localization of spin $1/2$ fermion fields is also interesting. It has been proved that, in order to normalize the zero mode, the fermion fields should couple with the background scalars. With different scalar-fermion couplings, there will exist a single bound state and a continuous gapless spectrum of massive fermion Kaluza-Klein (KK) states \cite{1006.4240,Liu0708,20082009,Fermions2010,Castro,zhao2011jhep}, or exist finite discrete KK states (mass gap) and a continuous gapless spectrum starting at a positive $m^2$ \cite{SingletonFermions,Liu0803,zhaoliu2010,George,fuGRS,0803.1458}, or even only exist bound KK modes \cite{Liu0907.0910,Liu6,guo2011}. We will show that with one type of scalar-fermion coupling there also can exist the above three cases, which is decided by {the value of the dilaton-fermion coupling constant}.

This paper is organized as follows: In Sec. \ref{SecModel}, we use a numerical method to obtain a solution of a thick flat split-brane, in which the extra dimension is finite. Then, in Sec. \ref{SecLocalize}, we study the stability of the solution and investigate localization of vector, KR, and fermion fields on this brane world. Finally, a brief discussion and conclusion are given in Sec. \ref{secConclusion}.

\section{The deformed brane with finite extra dimension}
\label{SecModel}

In this paper we consider the model of thick branes generated by two interacting scalars $\phi$ and $\pi$. The corresponding action of the system is
\begin{equation}
 \label{action}
 S=\!\!\int\!\! d^{5}x\sqrt{-g}\bigg[\frac{R}{2\kappa_5^2}-\frac{1}{2}(\partial\phi)^{2}
   -\frac{1}{2}(\partial\pi)^{2}-\bar{V}(\phi,\pi)\bigg],
\end{equation}
where $\kappa_5^2=8 \pi G_5$ with $G_5$ the 5-dimensional Newton constant, and $\bar{V}(\phi,\pi)$ is the potential making the thick branes be realized naturally. Here we set $\kappa_5=1$. The line-element of the 5-dimensional space-time is assumed as
\begin{eqnarray}
 ds^2&=&\text{e}^{2A(z)}\big(\eta_{\mu\nu}dx^\mu dx^\nu
          + dz^2\big),\label{line-element}
\end{eqnarray}
where $z$ stands for the extra coordinate, and $\text{e}^{2A(z)}$ is the warp factor. According to the symmetry of the metric, we can suppose that the background scalars $\phi, \pi$ are only the functions of $z$. Then the equations of motion from the action (\ref{action}) with the ansatz (\ref{line-element}) read:
\begin{eqnarray}\label{EOM1}
 \phi'^2 + \pi'^2 &=& 3(A'^2-A''), \\
 \label{EOM2}
 -2\;\bar{V}(\phi,\pi) &=& 3\;\text{e}^{-2A}(3A'^2+A''),\\
 \label{EOM3}
 \frac{\partial \bar{V}(\phi,\pi)}{\partial \phi}&=&\text{e}^{-2A}\big(\phi'' + 3A'\phi' \big),\\
 \label{EOM4}
 \frac{\partial \bar{V}(\phi,\pi)}{\partial \pi}&=&\text{e}^{-2A}\big(\pi'' + 3A'\pi' \big).
\end{eqnarray}

Firstly we analyze the character of the extra dimension through  Eq.~(\ref{EOM1}). Inspired by the idea in Ref. \cite{two-scalarbrane1}, we assume that the dilaton field has the form $\pi=\sqrt{3b}\;A$ with $b$ a positive constant, thus Eq. (\ref{EOM1}) is reduced to
\begin{equation}
  A''=-\frac{1}{3}\phi'^2+(1-b)A'^2.
\end{equation}
So for $\phi'(z\rightarrow \pm\infty)\rightarrow0$, we have $A''=(1-b)\;A'^2$, from which we could obtain the {behavior} of the warp factor at infinity:
\begin{equation}
  A(z\rightarrow \pm\infty)\rightarrow
  \left\{
  \begin{array}{lc}
    \frac{\ln{[|(1-b)z|}]}{b-1} & ~\text{for}~ b \neq 1 \\
    k|z| & ~\text{for}~ b=1
  \end{array}
  \right.  .
\end{equation}
Here we are interested in the case of $b\neq1$.
Thus whether the physical length of the extra dimension $y$ is finite or not can be checked by performing a coordinate transformation $dy=\text{e}^{A}dz$. From the following expression
\begin{eqnarray}
  \int_{y_0}^{y_b}dy&=&\int_{z_0}^{+\infty}\text{e}^{A(z)}dz\nonumber \\
  &\rightarrow&  \int _{z_0}^{+\infty}
                 \big[|(1-b)|z\big]^{\frac{1}{b-1}}dz\nonumber \\
  &\propto& \frac{b-1}{b} \lim_{z\rightarrow+\infty}
                 {(z^{\frac{b}{b-1}}-z_0^{\frac{b}{b-1}})},
\end{eqnarray}
where $y_0$ and $z_0$ are positive constants, it is clear that, for $0<b<1$, we will get a finite extra dimension with two boundaries at $y=\pm y_b$. Therefore, throughout this paper, the parameter $b$ will be limited to $0<b<1$. Note that for the standard Randall-Sundrum brane scenario (the case $b=0$), it can be shown that the physical extra dimension is infinite.

For our choice of $\pi=\sqrt{3b}\;A$, we get the following relation from Eqs. (\ref{EOM2}) and (\ref{EOM4}) for the potential $\bar{V}(\phi,\pi)$:
\begin{equation}
  -2\bar{V}(\phi,\pi)=\sqrt{3/b}\;\frac{\partial\bar{V}(\phi,\pi)}{\partial\pi},
\end{equation}
which {results that} the potential should be taken the form of $\bar{V}=V(\phi)\text{e}^{-2\sqrt{b/3}\;\pi}$. And the equations of motion (\ref{EOM1})-(\ref{EOM4}) are simplified as:
\begin{eqnarray}
 A'' &=& -\frac{1}{3}\phi'^2+(1-b)A'^2, \label{eom21}\\
 \phi'' &=& \text{e}^{2(1-b)\;A}\frac{d V(\phi)}{d \phi}-3A'\phi',\label{eom22}\\
 2V(\phi) &=& \text{e}^{2(b-1)\;A}\big[\phi'^2+3(b-4)A'^2\big].\label{eom23}
\end{eqnarray}
Now there are left three equations for three variables, but these equations are not independent. So if given the potential
\begin{equation}
  V(\phi)=v_0+g_1 \phi^2-g_2\phi^4+g_3\phi^6
\end{equation}
with $v_0,g_1,g_2,g_3$ all positive constants, we can get the brane solutions by numerical method with the boundary conditions
\begin{eqnarray}
  A(0)=A'(0)=\phi(0)=0, \phi(z\rightarrow \pm \infty)=\pm\phi_0.\label{condition}
\end{eqnarray}
Here $\phi_0$ is a positive constant. For this brane world, it is the form of the dilaton that is crucial to the finity of the extra dimension.

\begin{figure*}[htb]
\begin{center}
\includegraphics[width=7.7cm]{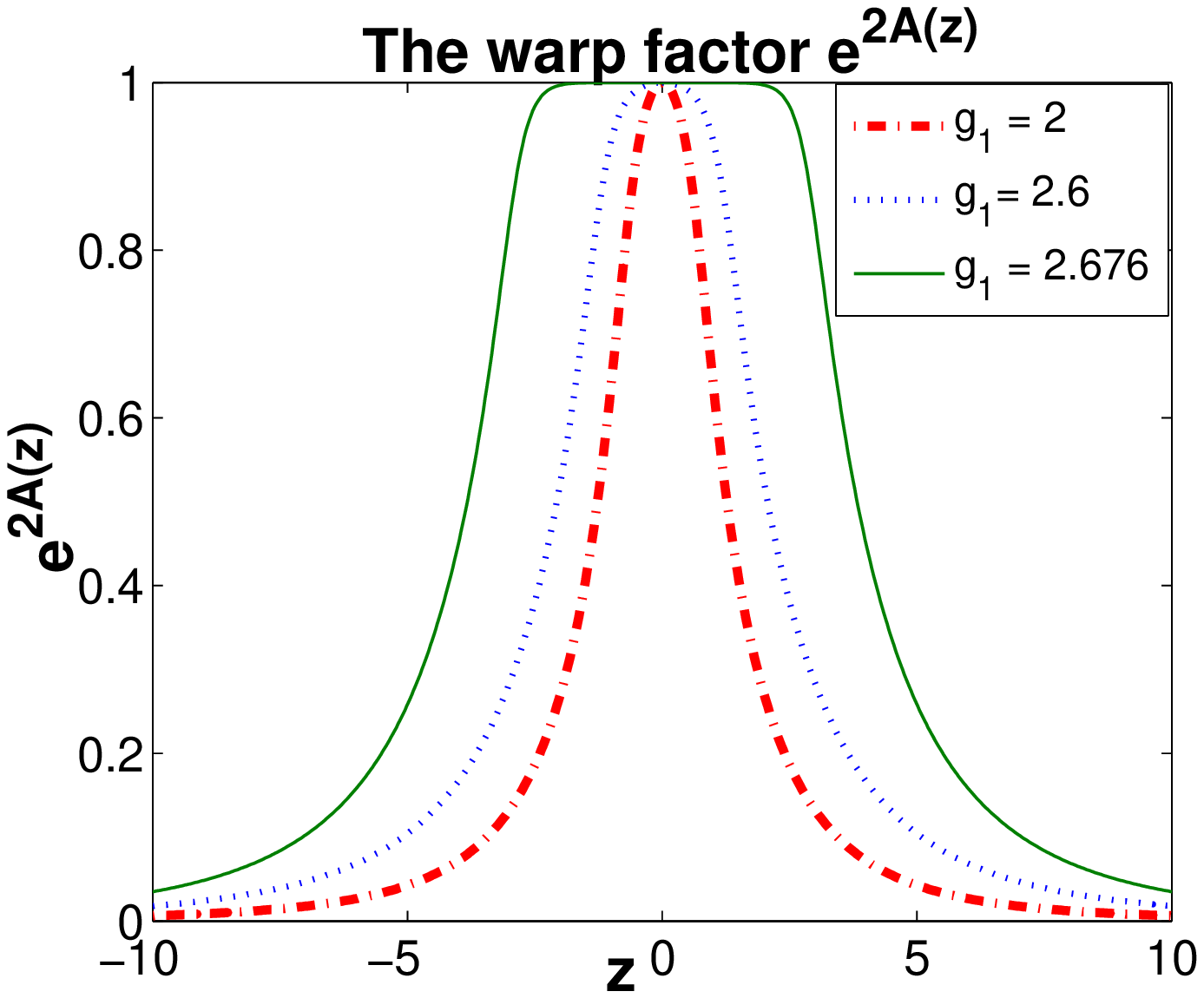}
\includegraphics[width=7.7cm]{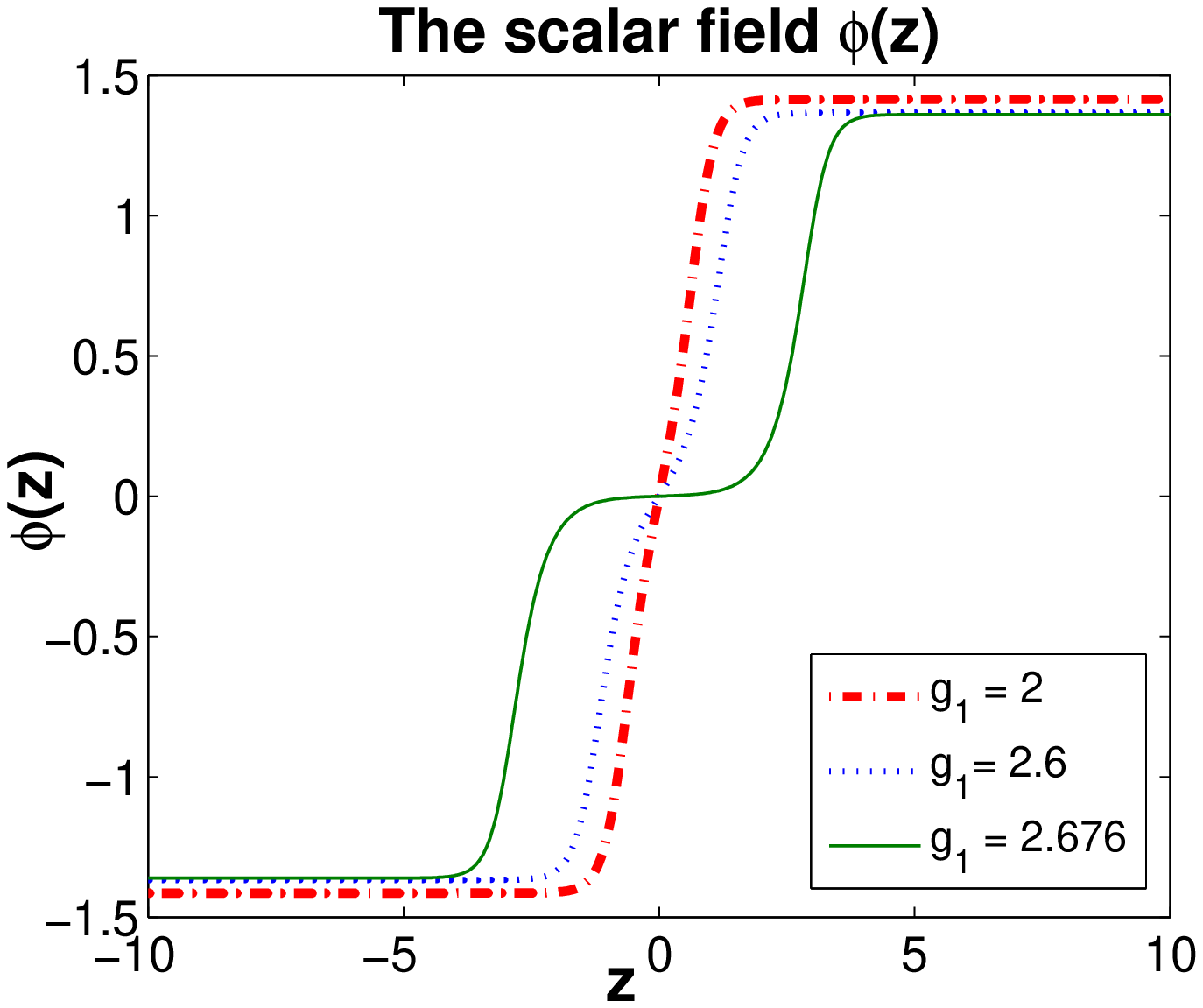}
\includegraphics[width=7.7cm]{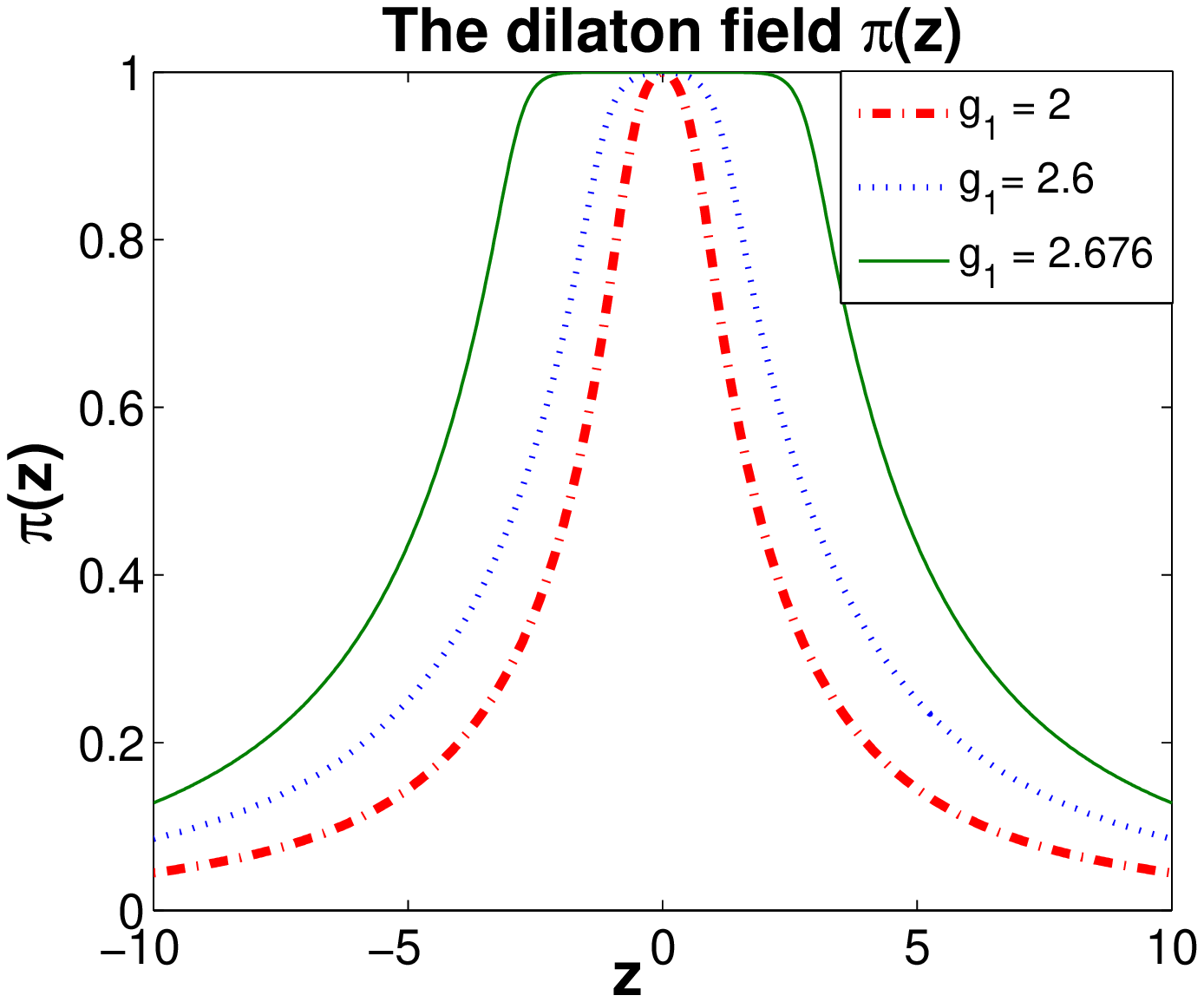}
\includegraphics[width=7.7cm]{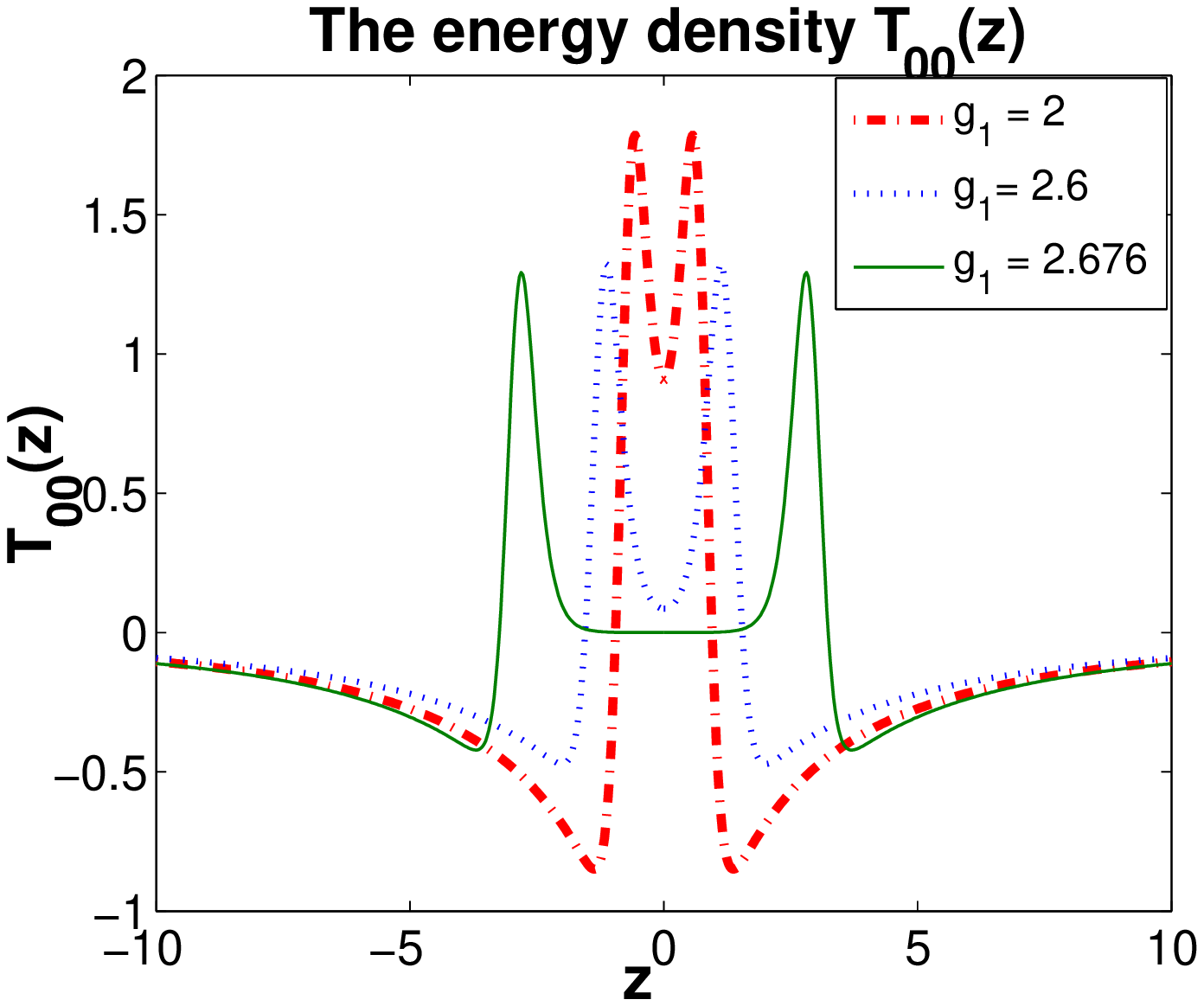}
\end{center}
\caption{The shapes of the warp factor $\text{e}^{2A(z)}$, the scalar field $\phi(z)$, the dilaton field $\pi(z)$ and the energy density $T_{00}(z)$. The parameters are set to $g_2=3.5,g_3=1,b=0.5$.}
 \label{figSolz}
\end{figure*}

The shapes of the warp factor, scalar fields $\phi$ and $\pi$, and the profiles of energy density $T_{00}$, which is defined as $T_{00}=\phi'^2+3(b-2)A'^2$, are plotted in Fig.~\ref{figSolz}. And we find that for some proper values of $g_1$ there exists double-kink solution for the scalar field $\phi$. Thus the brane splits into two sub-branes. With the increase of $g_1$, the distance of the two sub-branes will increase.

In the following section, we will investigate the localization of bulk matters in this braneworld. And we will mainly analyse the effect of the finity of the extra dimension and the dilaton scalar on the localization.

\section{Localization and mass spectra of various bulk matter fields on the brane}
\label{SecLocalize}

In this section we will investigate the localization and mass spectra of various bulk matter fields.

\begin{figure*}[htb]
\begin{center}
\includegraphics[width=8cm]{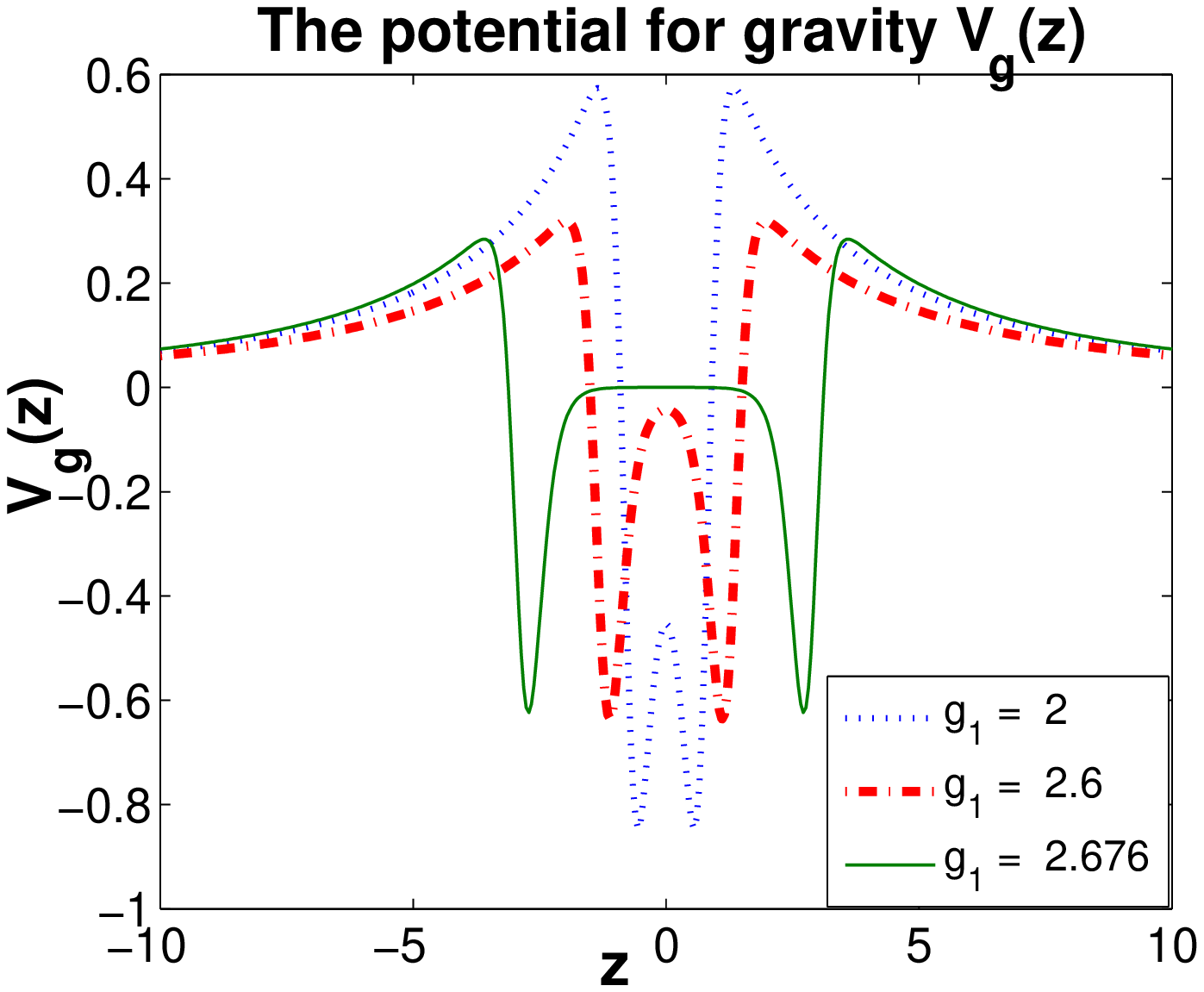}
\includegraphics[width=8cm]{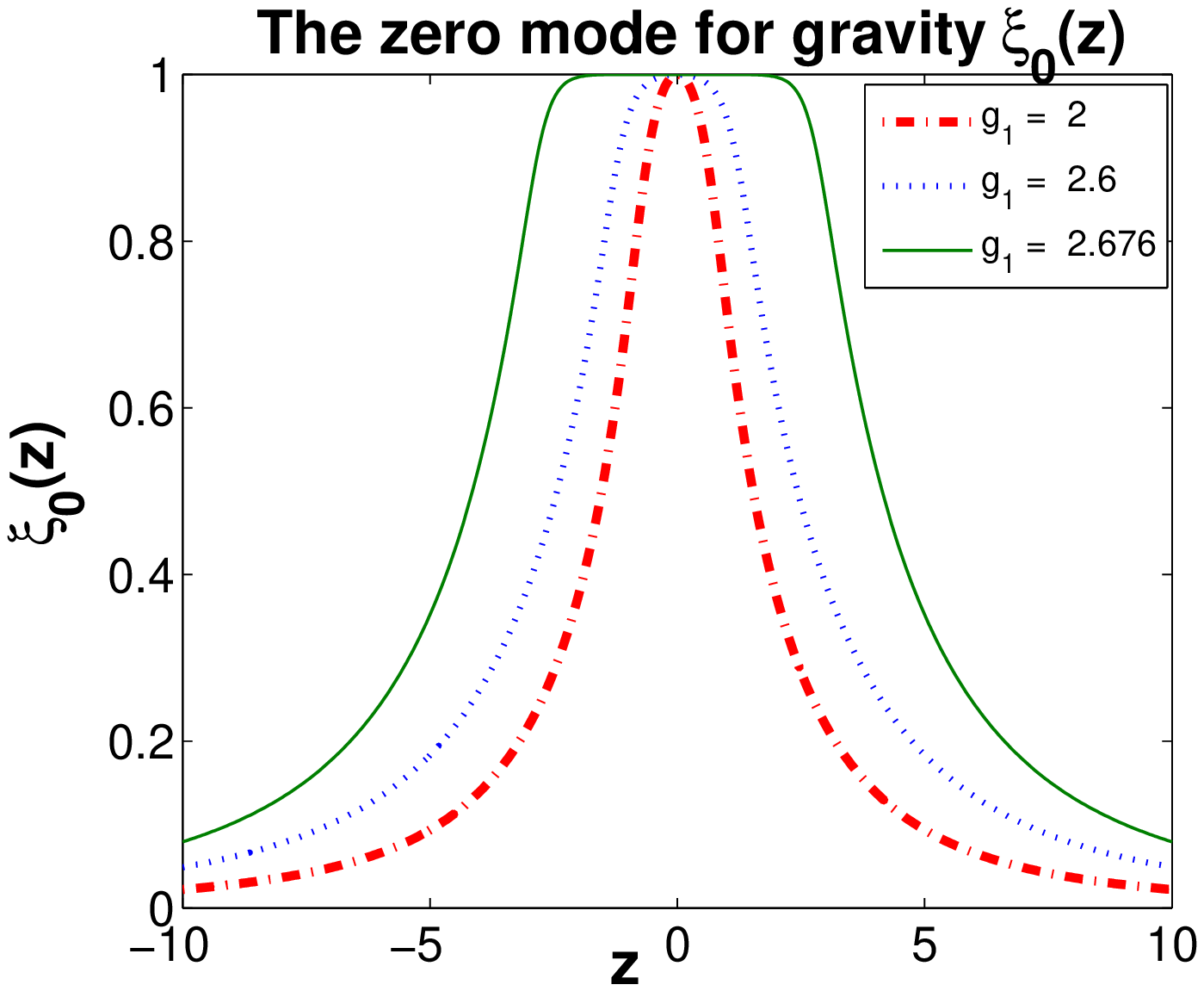}
\end{center}
\caption{The shapes of the potential $V_g(z)$ for the gravity KK modes and the gravity zero mode $\xi_0(z)$. The parameters are set to $g_2=3.5,g_3=1,b=0.5$.}
 \label{VG}
\end{figure*}

Firstly, we can see whether the solution is stable {by giving} the metric a fluctuation $h_{\mu\nu}$, so the metric (\ref{line-element}) reads as
\begin{equation}
  ds^2=\text{e}^{2A(z)}\big[(\eta_{\mu\nu}+h_{\mu\nu})dx^{\mu}dx^{\nu}
  +dz^2\big].
\end{equation}
Using the gauge choice $h_{ \mu}^\mu=\partial^\mu h_{\mu\nu}=0$, we can find the $h_{\mu\nu}$ takes the following form
\begin{equation}
  \big[\partial_z^2+3A'(z)\partial_z+\square^{(4)}\big]h_{\mu\nu}(x^a,z)=0
\end{equation}
with $\square^{(4)}\equiv\eta^{\mu\nu}\triangledown_\mu\triangledown_\nu$ and $\triangledown_\mu$ the covariant derivative with respect to the four-dimensional metric $\eta_{\mu\nu}$. Then through the KK decomposition $h_{\mu\nu}(x^a,z)=\epsilon(x^a)\xi(z)\text{e}^{-3A/2}$, we can obtain the following Schr\"{o}dinger equation for the KK modes
\begin{eqnarray}
  (-\partial_z^2+V_g)\xi(z)=m^2\xi(z),  \label{SchVG}
\end{eqnarray}
where $m^2$ are the masses of the KK modes and the effective potential is
$V_g=\frac{3}{2}A''+\frac{9}{4}A'^2$. Because the formulation (\ref{SchVG}) can be written as
\begin{equation}
  \Big[-\partial_z-\frac{3}{2}A'(z)\Big]\Big[\partial_z-\frac{3}{2}A'(z)\Big]
  \xi(z)
  =m^2\xi(z),
\end{equation}
there is no tachyon. We plot the shapes of the potential in Fig.~\ref{VG}, from which it can be seen that there exists a zero mode $\xi_0\varpropto\text{e}^{3A/2}$. As
\begin{equation}
   \xi_0^2\rightarrow \frac{1}{\big[(1-b)z\big]^{\frac{3}{2(1-b)}}}\rightarrow0,
\end{equation}
the integral $\int \xi_0^2 dz$ is finite for $0<b<1$,
so the zero mode (4-dimensional massless graviton) can be localized on the brane.
The shape of the zero mode is also plotted in Fig.~\ref{VG}.

\subsection{Spin-1 vector fields}

Now we investigate {the localization of} the massless spin-1 vector fields in 5-dimensional space-time. The action of a massless vector field coupled with gravity and the dilaton is
\begin{eqnarray}
S_1 = - \frac{1}{4} \int d^5 x
\sqrt{-g}\;\text{e}^{\sigma\pi} g^{M R} g^{N S} F_{MN}
F_{RS}, \label{actionVector}
\end{eqnarray}
where the field strength tensor is given by $F_{MN} = \partial_M A_N - \partial_N A_M$ and $\sigma$ is the coupling constant. Then the equations of motion can be obtained using the background geometry (\ref{line-element}):
\begin{eqnarray}
\text{e}^{\sigma\pi}\frac{1}{\sqrt{-\hat{g}}}\partial_\nu (\sqrt{-\hat{g}}
      {\hat{g}}^{\nu \rho} \hat{g}^{\mu\lambda}F_{\rho\lambda})&&\nonumber \\
    +{\hat{g}^{\mu\lambda}}\text{e}^{-A}\partial_z
      \left(\text{e}^{(A+\sigma\pi)} F_{z\lambda}\right)  &=& 0, \\
 \text{e}^{\sigma\pi}\partial_\mu(\sqrt{-\hat{g}}\hat{g}^{\mu\nu}F_{\nu 4})&=& 0.
\end{eqnarray}
With the decomposition of the vector field $A_{\mu}(x,z)=\sum_n a^{(n)}_\mu(x)\rho_n(z)\text{e}^{-(1+\sqrt{3b}\;\sigma)A/2}$ and the gauge choice $A_4=0$, we find that the KK modes of the vector field satisfy the following Schr\"{o}dinger-like equation:
\begin{eqnarray}
  \left[-\partial^2_z +V_1(z) \right]{\rho}_n(z)=m_n^2
  {\rho}_n(z)  \label{SchEqVector1}
\end{eqnarray}
with $m_n$ the masses of the 4-dimensional vectors, and
\begin{eqnarray}
  V_1(z)=\frac{(1+\sqrt{3b}\;\sigma)^2}{4} A'^2
         +\frac{1+\sqrt{3b}\;\sigma}{2}  A''. \label{V1}
\end{eqnarray}
And furthermore providing the orthonormality {condition}
\begin{eqnarray}
 \int^{\infty}_{-\infty} dz \;\rho_m(z)\rho_n(z)=\delta_{mn},
 \label{normalizationConditionVecter}
\end{eqnarray}
we can get the 4-dimensional effective action:
\begin{eqnarray}
 S_1 = \sum_{n}\int d^4 x \sqrt{-\hat{g}}~
       \bigg( \!\!&-& \!\! \frac{1}{4}\hat{g}^{\mu\alpha} \hat{g}^{\nu\beta}
             f^{(n)}_{\mu\nu}f^{(n)}_{\alpha\beta}
                \nonumber \\
     &-& \!\! \frac{1}{2}m^2_{n} ~\hat{g}^{\mu\nu}
           a^{(n)}_{\mu}a^{(n)}_{\nu}
       \bigg),
\label{actionVector2}
\end{eqnarray}
where $f^{(n)}_{\mu\nu} = \partial_\mu a^{(n)}_\nu -
\partial_\nu a^{(n)}_\mu$ is the 4-dimensional field strength
tensor.

\begin{figure*}[htb]
\begin{center}
\includegraphics[width=8.1cm]{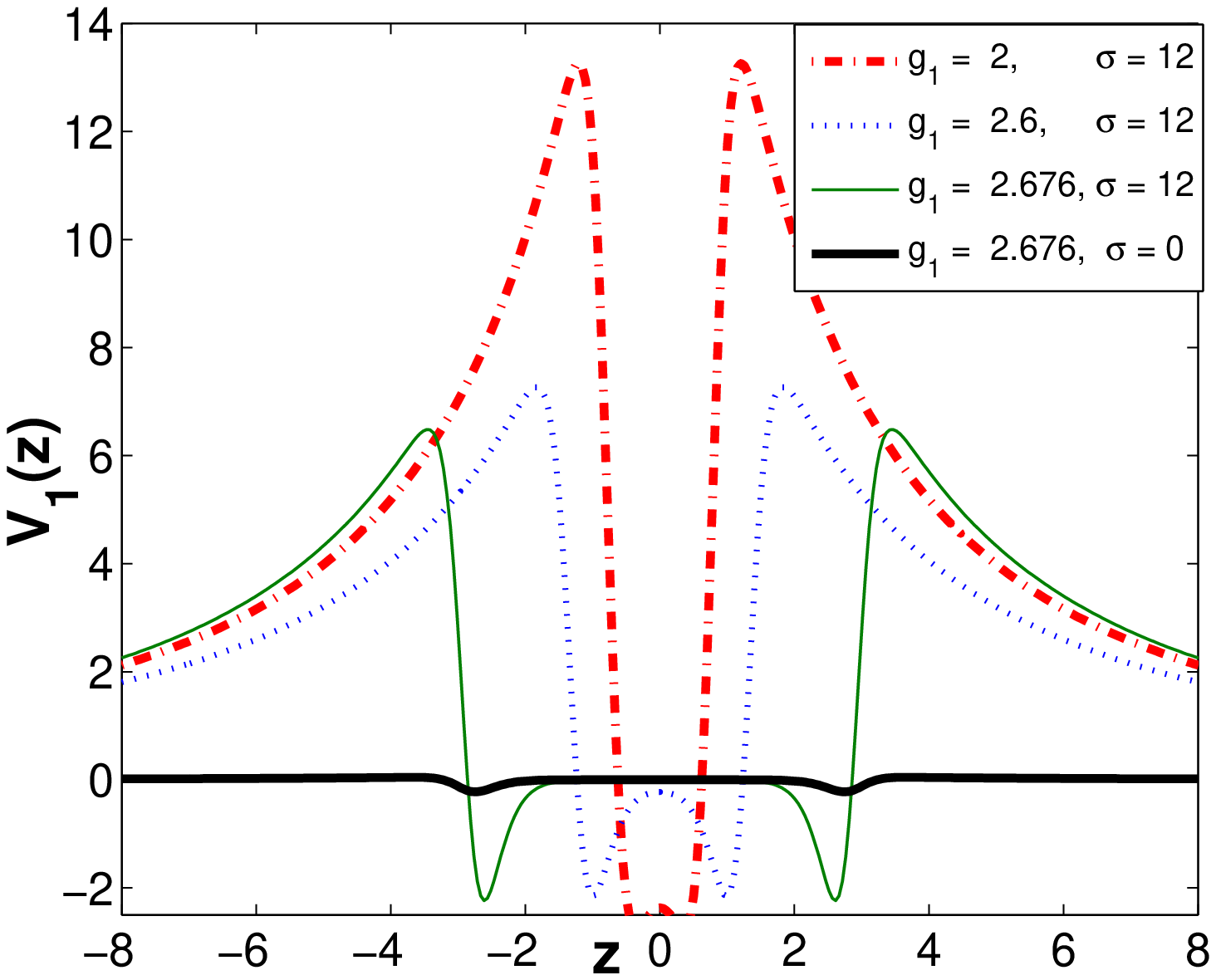}
\includegraphics[width=8.1cm]{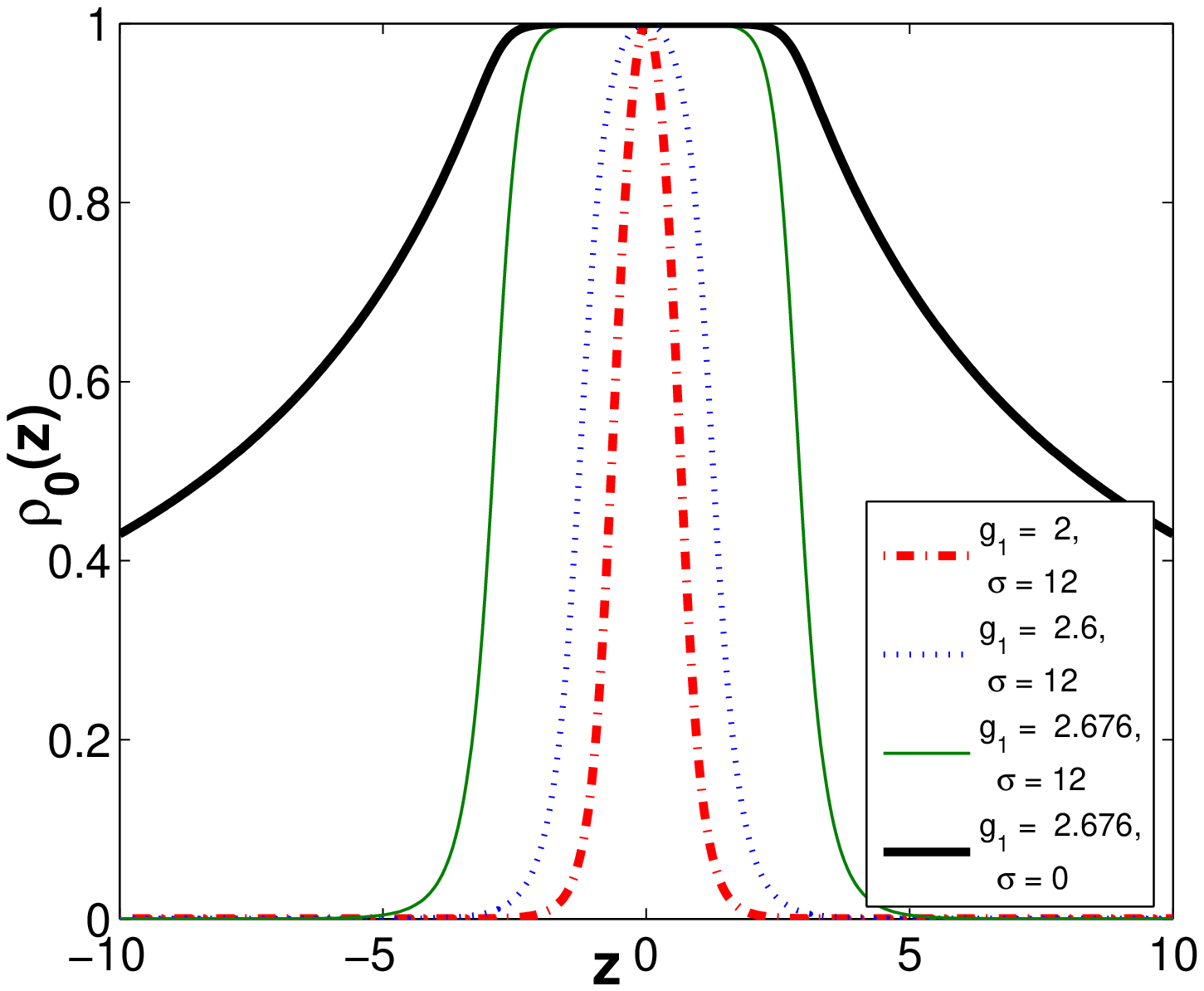}
\end{center}
 \caption{The shapes of the potential $V_1(z)$ and the vector zero mode $\rho_0(z)$ with $g_2=3.5,g_3=1,b=0.5$.}
 \label{FigVector}
\end{figure*}

With the conditions (\ref{condition}), it can be found that the potential trends to $\frac{-1-\sqrt{3b}\;\sigma}{6}\phi'(0)^2$ at $z=0$, and vanishes at infinity. So when $\sigma>-1/\sqrt{3b}$, there may exist a bound zero mode, which can be obtained by setting $m_0=0$:
\begin{eqnarray}
\rho_0\varpropto \text{e}^{\frac{(1+\sqrt{3b}\;\sigma)}{2}A(z)}.
\end{eqnarray}
We can check whether it satisfies the normalization condition (\ref{normalizationConditionVecter}):
\begin{eqnarray}
  \int^{\infty}_{-\infty} \;\rho_0^2 dz
  \varpropto
  \int^{\infty}_{-\infty} \;\text{e}^{{(1+\sqrt{3b}\;\sigma)}A(z)} dz
< \infty,
\end{eqnarray}
which is equivalent to
\begin{eqnarray}
  \int^{\infty}_{1} \big[(1-b)z\big]^{\frac{1+\sqrt{3b}\;\sigma}{b-1}}dz < \infty.
\end{eqnarray}
We can find that the vector zero mode can be localized on the region between the sub-branes with $\sigma>-\sqrt{b/3}$ and $0<b<1$. If there is no coupling ($\sigma=0$), the vector zero mode will be $\rho_0=c_0\text{e}^{A/2}$, so with the coordinate transformation $dz=\text{e}^{-A}dy$ the integral (\ref{normalizationConditionVecter}) becomes
 \begin{eqnarray}
   \int^{\infty}_{-\infty} \rho_0^2 dz=\int^{y_b}_{-y_b} c_0^2dy=2c_0^2 y_b=1,
 \end{eqnarray}
and the normalization coefficient is $c_0=1/\sqrt{2y_b}$.
Because the physical extra dimension $y$ is finite, the above integral is finite.
Hence, the vector zero mode can also be localized on the brane even without the coupling, which could be seen from Fig.~\ref{FigVector}.
The shapes of the potential with the coupling and the zero mode are also plotted in Fig.~\ref{FigVector}.

From Fig.~\ref{FigVector}, we can see that there may exist some vector resonances for $\sigma\neq0$. Following Refs.~\cite{LiuYang2009}
we calculate numerically the masses and the life-time
for the resonances in Table~\ref{tab1},
and find that the number and the lift-time of the resonances increase with the distance between the two sub-branes.

\begin{table}[h]
\begin{center}
\renewcommand\arraystretch{1.1}
\begin{tabular}
 {|l|c|c|c|c|c|c|}
  \hline
  $~~g_1$ & $n$ & $m^2$ & m & $\delta m$ & $\tau$  \\
  \hline \hline
   ~~2   &  $1$ & 6.63921 & 2.57667 & 0.006518451883 & 153.41066 \\
  \hline \hline
   ~2.6  & $1$ & 1.94348 & 1.39409 & 0.000118884093 & 8411.55427 \\
       \cline{2-6}
        & $2$ & 5.69137 & 2.38566 & 0.078850635906 & 12.68221\\
  \hline \hline
         & $1$ & 0.33613 & 0.57976 & 1.9641206011$\times10^{-5}$ & 50913.37057\\
  \cline{2-6}
   2.676 &  $2$ & 1.30244 & 1.14124 & 7.3715986182$\times10^{-5}$ & 13565.57854	\\
  \cline{2-6}
        &  $3$ & 2.79574 & 1.67205 & 0.002230625464 & 448.30475\\
  \cline{2-6}
        &  $4$ & 4.65884 & 2.15844 & 0.038868015350 & 25.72810 \\
  \hline
\end{tabular}
\end{center}
\caption{The mass $m^2$, width $\delta m$, and life-time $\tau$ for resonances of the vector field for different $g_1$.}\label{tab1}
\end{table}

\subsection{The Kalb-Ramond fields}

We {now turn to} the KR fields in this subsection. The action of a KR field coupled with the background dilaton scalar $\pi$ is
\begin{equation}
S_{\text{KR}} = -\int d^{5}x \sqrt{-g}~ \text{e}^{\zeta \pi} H_{MNL}H^{MNL}\label{actionKRPhi}
\end{equation}
with $H_{MNL}=\frac{1}{6}\partial_{[M}B_{NL]}$ the field strength for the KR field $B_{MN}$ and $\zeta$ the coupling constant between the KR field and the dilaton field. Then the equations of motion derived from this action and the conformally flat metric (\ref{line-element}) are:
\begin{eqnarray}
 \text{e}^{\zeta\pi}\partial_\mu ( \sqrt{-g}H^{\mu\alpha\beta})
 +\partial_4(\sqrt{-g}\text{e}^{\zeta\pi}H^{4\alpha\beta})&=& 0, \\
 \text{e}^{\zeta\pi}\partial_\mu ( \sqrt{-g}H^{\mu4\beta})&=& 0.
\end{eqnarray}
If we choose the gauge $B_{\alpha4}=0$ and make a decomposition of the KR field as $B^{\alpha\beta}_{(n)}(x^\lambda,z)=\sum_n \hat{b}^{\alpha\beta}_{(n)}(x^\lambda)U_{n}(z)\text{e}^{
(-7-\sqrt{3b}\;\zeta)A/2}$, we can get the Schr\"{o}dinger equation for $U_{n}(z)$:
\begin{eqnarray}
\big[ -\partial^2_z+ V_{\text{KR}}(z)\big]U_n(z)
  =m_n^2 U_n(z),
  \label{SchEqK-R}
\end{eqnarray}
where $m_n$ are the masses of the 4-dimensional KR fields and the effective potential is
\begin{eqnarray}
 V_{\text{KR}}=\frac{(\sqrt{3b}\;\zeta-1)^2}{4}A'^2
+\frac{\sqrt{3b}\;\zeta-1}{2}A''.
\end{eqnarray}
Provided the orthonormality {condition} $\int^{\infty}_{-\infty} dz \;U_m(z)U_n(z)=\delta_{mn}$, the action of the KR field (\ref{actionKRPhi}) is reduced to the following 4-dimensional effective action:
\begin{eqnarray}
 S_{\text{KR}} = -\sum_{n}\int d^4 x \sqrt{-\hat{g}}~
       \bigg( \hat{g}^{\mu'\mu}\hat{g}^{\alpha'\alpha}\hat{g}^{\beta'\beta}
       \hat{h}_{\mu'\alpha'\beta'}^{(n)}\hat{h}_{\mu\alpha\beta}^{(n)}
       \nonumber \\
       +\frac{1}{3}m_n^2\hat{g}^{\alpha'\alpha}\hat{g}^{\beta'\beta}
       \hat{b}_{\alpha'\beta'}^{(n)}\hat{b}_{\alpha\beta}^{(n)}
       \bigg)
\label{actionKR2}
\end{eqnarray}
with $\hat{h}_{\mu\alpha\beta}^{(n)}=\partial_{[\mu}\hat{b}_{\alpha\beta]}$ the 4-dimensional field strength tensors.

We can find the behavior of the potential for the KR field at $z=0$ and $z\rightarrow \infty$ :
\begin{eqnarray}
  V_{\text{KR}}(z\rightarrow0)&\rightarrow&\frac{1-\sqrt{3b}\;\zeta}{6}\phi'(0)^2,\\
  V_{\text{KR}}(z\rightarrow\infty)&\rightarrow&0.
\end{eqnarray}
So in order to localize the zero mode, we must make sure $\zeta>1/\sqrt{3b}$.

By setting $m_0=0$, we can solve the zero mode for the KR field:
\begin{eqnarray}
  U_{0}\propto\text{e}^{\frac{\sqrt{3b}\;\zeta-1}{2} A}.
\label{KRzeromode}
\end{eqnarray}
One can check whether the zero mode satisfies the normality condition $\int^{\infty}_{-\infty} dz \;U^2_0(z)<\infty$. {Because} we have
\begin{eqnarray}\label{KRzeromode}
  \int U_{0}^2 dz \propto
  \big[(1-b)|z|\big]^{\frac{\sqrt{3b}\;\zeta-2+b}{b-1}} ~~\text{for}~~ z\rightarrow \pm\infty,
\end{eqnarray}
for $\zeta>(2-b)/\sqrt{3b}$ the zero mode can be localized on the brane, but it can not for $\zeta=0$, which is different with the vector field.

\subsection{The spin-1/2 fermion fields}

In this subsection, we investigate the spin-1/2 fermion fields. The Dirac action of a massless spin $1/2$ fermion coupled with gravity and the background scalars $\phi$ and $\pi$ in 5-dimensional space-time is
\begin{eqnarray}
 S_{\frac{1}{2}} = \! \int \! d^5 x \sqrt{-g} \Big[\bar{\Psi} \Gamma^M (\partial_M + \omega_M) \Psi-\eta \bar{\Psi} F(\phi, \pi) \Psi\Big]
 \label{DiracAction}
\end{eqnarray}
with $F(\phi, \pi)$ the type of the coupling and $\eta$ the coupling constant. {As in Refs.~\cite{LiuYang2009}}, we have $\Gamma^M=(\text{e}^{-A}\gamma^{\mu},\text{e}^{-A}\gamma^5)$,
$\omega_\mu =\frac{1}{2}(\partial_{z}A) \gamma_\mu \gamma_5$,
$\omega_5 =0$, where $\gamma^{\mu}$ and $\gamma^5$ are the usual flat gamma matrices in the 4-dimensional Dirac representation. Then the 5-dimensional Dirac equation is
\begin{eqnarray}
 \left[ \gamma^{\mu}\partial_{\mu}
         + \gamma^5 \left(\partial_y  +2 \partial_{z} A \right)
         -\eta¡¡\text{e}^A F(\phi, \pi)
 \right] \Psi =0, \label{DiracEq1}
\end{eqnarray}
where $\gamma^{\mu}\partial_{\mu}$ is the 4-dimensional Dirac operator. By the general chiral decomposition $ \Psi(x,z) = \text{e}^{-2A}\sum_n\big(\psi_{Ln}(x) f_{Ln}(z)
 +\psi_{Rn}(x) f_{Rn}(z)\big)$, we get the following coupled equations of $f_{Ln}(z)$ and $f_{Rn}(z)$:
\begin{subequations}\label{CoupleEq1}
\begin{eqnarray}
 \left[\partial_z + \eta\;\text{e}^A F(\phi, \pi) \right]f_{Ln}(z)
  &=&  ~~m_n f_{Rn}(z), \label{CoupleEq1a}  \\
 \left[\partial_z- \eta\;\text{e}^A F(\phi, \pi) \right]f_{Rn}(z)
  &=&  -m_n f_{Ln}(z), \label{CoupleEq1b}
\end{eqnarray}
\end{subequations}
where $\psi_{Ln,Rn}(x)$ satisfy the four-dimensional massive Dirac equations $\gamma^{\mu}\partial_{\mu}\psi_{Ln}(x)
=m_n\psi_{Rn}(x)$ and $\gamma^{\mu}\partial_{\mu}\psi_{Rn}(x)
=m_n\psi_{Ln}(x)$. Furthermore we can get the following Schr\"{o}dinger-like equations for the KK modes of the left- and right-hand fermions from the above coupled equations:
\begin{subequations}\label{SchEqFermion}
\begin{eqnarray}
  \big(-\partial^2_z + V_L(z) \big)f_{Ln}
            &=&m_n^2 f_{Ln},
   \label{SchEqLeftFermion}  \\
  \big(-\partial^2_z + V_R(z) \big)f_{Rn}
            &=&m_n^2 f_{Rn},
   \label{SchEqRightFermion}
\end{eqnarray}
\end{subequations}
where the effective potentials take the following form
\begin{subequations}
\begin{eqnarray}
  V_L(z)&=& \big(\eta\;\text{e}^{A}   F(\phi, \pi)\big)^2
     - \eta\partial_z \big(\;\text{e}^{A}   F(\phi,\pi)\big), \label{VL}\\
  V_R(z)&=&   V_L(z)|_{\eta \rightarrow -\eta}. \label{VR}
\end{eqnarray}\label{Vfermion}
\end{subequations}
And we can obtain the standard 4-dimensional action
for a massless and a series of massive fermions:
\begin{eqnarray}
 S_{\frac{1}{2}}=\sum_n\int d^4x \bar{\psi}_{n}(x)
 \big[\gamma^\mu \partial_\mu-m_n\big]\psi_{n}(x)
\end{eqnarray}
with the following orthonormality conditions for
$f_{L{n}}$ and $f_{R{n}}$:
\begin{eqnarray}
 \int_{-\infty}^{\infty} f_{Lm} f_{Ln}dz
  &=&\delta_{mn},\nonumber\\
  \int_{-\infty}^{\infty} f_{Rm} f_{Rn}dz
  &=&\delta_{mn},\\
  \int_{-\infty}^{\infty} f_{Lm} f_{Rn}dz &=& 0.\nonumber
\end{eqnarray}

If {the type of the scalar-fermion coupling is considered to be} $F(\phi, \pi)=\text{e}^{\lambda \pi}\phi$ with $\lambda$ the dilaton-fermion coupling constant, we can get
\begin{subequations}
\begin{eqnarray}
  V_L(z)&=&\eta \text{e}^{(1+\sqrt{3b}\lambda)A}
        \bigg(\eta\phi^2\text{e}^{(1+\sqrt{3b}\lambda)A}
        \nonumber \\
        && -(1+\sqrt{3b}\lambda)A'\phi-\partial_z\phi \bigg), \label{VL2}\\
  V_R(z)&=& V_L(z)|_{\eta \rightarrow -\eta}. \label{VR2}
\end{eqnarray}\label{Vfermion2}
\end{subequations}
Thus we obtain the behavior of the potential at $z=0$ and $z\rightarrow \infty$:
\begin{subequations}\label{VfermionInfinity}
\begin{eqnarray}
  V_{L,R}(z\rightarrow0)\!\!&=&\!\!\eta \phi'(0),\\
  V_L(z\rightarrow\pm\infty)\!\!&=&\!\!\eta^2 \phi_0^2 \big[(1-b)|z|\big]^{\frac{2(1+\sqrt{3b}\lambda)}{b-1}}
  \nonumber \\
    \!\!&-&\!\! \eta \phi_0(1\!\!+\!\!\sqrt{3b}\lambda) \big[(1\!\!-\!\!b)|z|\big]^{\frac{(1\!+\!\sqrt{3b}\lambda)}{b-1}-\!1}, ~~~~~\label{VL2}\\
  V_R(z\rightarrow\pm\infty)\!\!&=&\!\! V_L(z\pm\infty)|_{\eta \rightarrow -\eta}. \label{VR2}
\end{eqnarray}
\end{subequations}
 So it can be seen that the behavior of the potentials at $z\rightarrow\pm\infty$ is decided by the dilaton-fermion coupling constant $\lambda$, and there are three types of potentials for different $\lambda$, which are similar with that in Ref.~\cite{branep1}. For $\lambda=-1/\sqrt{3b}$ and $\lambda>-1/\sqrt{3b}$ the potential for left-hand fermion is PT-like and volcano-like ones, respectively. For $\lambda<-1/\sqrt{3b}$ it is infinite potential well. And there is always a zero mode for the left-hand fermion with $\eta>0$ for these three cases, which takes the form:
  \begin{equation}
 f_{L0}(z) \propto \exp\left(-\eta\int^z_0 d\overline{z}
 \text{e}^{A(\overline{z})}F(\phi(\overline{z}), \pi(\overline{z}))\right).
  \label{zeroModefL0}
\end{equation}
Then we can check whether the zero mode can be localized on the brane by checking whether the following integral
\begin{equation}
 \int f_{L0}^2(z) dz \propto
 \int \exp\left(-2\eta\int^z_0 d\overline{z}
 \text{e}^{A(\overline{z})}F(\phi(\overline{z}), \pi(\overline{z}))\right) dz
 \label{condition1}
\end{equation}
is finite. As we have
\begin{eqnarray}
 \;\text{e}^{A} F(\phi, \pi)\rightarrow \big[(1-b)z\big]^{\frac{1+\sqrt{3b}\lambda}{b-1}}~~ \text{when} ~~z \rightarrow \infty,
\end{eqnarray}
the integral (\ref{condition1}) becomes:
\begin{equation}
 \int f_{L0}^2(y) dy \rightarrow
 \int \text{e}^{-\frac{2\eta}{1-b}\big[(1-b)z\big]^{\frac{\sqrt{3b}\lambda
 +b}{b-1}}}dz~~ \text{when} ~~z \rightarrow \infty.
 \label{condition2}
\end{equation}
So it is clear that for $\lambda<-\sqrt{b/3}$ the integral is finite, and hence the zero mode can be localized on the brane.

\section{Discussions and conclusions}
\label{secConclusion}

In this paper, we obtain a solution of a thick flat split-braneworld, which is generated by a double-kink scalar and a dilaton scalar. In this brane model, the extra dimension is finite, which is due to the choice { of the form of the dilaton field $\pi=\sqrt{3b}\;A$ with $0<b<1$}. And {because} another scalar is a {double-kink} one, the brane is split.

Because the extra dimension is finite, so the zero mode for the spin-1 vector fields can be localized on the brane. And if we consider the coupling to the dilaton, there will be resonances. The number of the resonances will increase with the distance between the two sub-branes.

But for KR fields, we have to introduce the coupling to the dilaton to {localize} the zero mode. And only when the coupling constant satisfies $\zeta>(2-b)/\sqrt{3b}$, the zero mode can be localized.

For spin $1/2$ fermion fields, we considered the coupling both with the double-kink and the dilaton, i.e., $\eta\bar{\Psi}F(\phi, \pi)\Psi$ with $F(\phi, \pi)=\text{e}^{\lambda\pi}\phi$. It was shown that there are three types of potentials for both chiral fermions, which are decided by the value of the dilaton-fermion coupling constant $\lambda$. However, only when $\lambda<-\sqrt{b/3}$, the zero mode for the left-hand fermion can be localized on the brane with $\eta>0$.

\section{Acknowledgement}

We would like to thank the referee for his/her useful comments and suggestions, which are very helpful to improve this paper. This work was supported by the Program for New Century Excellent
Talents in University, the Huo Ying-Dong Education Foundation of
Chinese Ministry of Education (No. 121106), the National Natural
Science Foundation of China (No. 11075065), the Doctoral Program Foundation of
Institutions of Higher Education of China (No.
20090211110028), the Fundamental Research Funds for the Central Universities (No. lzujbky-2012-k30 and No.lzujbky-2012-207).

\end{document}